\documentclass[aps,pra,onecolumn,superscriptaddress]{revtex4}

\usepackage{epstopdf}
\usepackage{color}
\usepackage{url}
\usepackage{hyperref}
\usepackage[T1]{fontenc}
\usepackage[latin1]{inputenc}
\usepackage{float}

\usepackage{physics}
\usepackage{amssymb} 
\usepackage{graphicx} 

\usepackage[english]{babel} 
\usepackage{blindtext} 

\usepackage{graphicx}
\usepackage{dcolumn}
\usepackage{bm}
\usepackage{amsmath}
\usepackage{upgreek}
\usepackage{bm}
\usepackage{tabularx}
\newcommand{\loss}[1]{\mathcal{L}_{\mathrm{#1}}}
\newcommand{\leff}[0]{L_{\mathrm{eff}}}
\usepackage{siunitx}

\begin{document}

\title{Efficient operator method for modelling mode mixing in misaligned optical cavities}

\author{W. J. Hughes}
\email[email: ]{william.hughes@physics.ox.ac.uk}
\affiliation{Department of Physics, University of Oxford, Clarendon Laboratory, Parks Rd, Oxford, OX1 3PU, UK}
\author{T. H. Doherty}
\affiliation{Department of Physics, University of Oxford, Clarendon Laboratory, Parks Rd, Oxford, OX1 3PU, UK}
\author{J. A. Blackmore}
\affiliation{Department of Physics, University of Oxford, Clarendon Laboratory, Parks Rd, Oxford, OX1 3PU, UK}
\author{P. Horak}
\affiliation{Optoelectronics Research Centre, University of Southampton, Southampton SO17 1BJ, UK}
\author{J. F. Goodwin}
\email[email: ]{joseph.goodwin@physics.ox.ac.uk}
\affiliation{Department of Physics, University of Oxford, Clarendon Laboratory, Parks Rd, Oxford, OX1 3PU, UK}

\date{\today}

\begin{abstract}
The transverse field structure and diffraction loss of the resonant modes of Fabry-P\'erot optical cavities are acutely sensitive to the alignment and shape of the mirror substrates. We develop extensions to the `mode mixing' method applicable to arbitrary mirror shapes, which both facilitate fast calculation of the modes of cavities with transversely misaligned mirrors and enable the determination and transformation of the geometric properties of these modes. We show how these methods extend previous capabilities by including the practically-motivated case of transverse mirror misalignment, unveiling rich and complex structure of the resonant modes.
\end{abstract}

\maketitle
\section{Introduction}
\label{sec: introduction}
The majority of Fabry-P\'erot optical cavities have mirrors with sufficiently constant curvature to be described well by standard resonator theory \cite{Siegman:86}. However, there are applications of cavities with non-spherical mirrors for which standard theory is not suitable. As a first example, the desire to realise stronger light matter coupling, whether to increase the rate of single photon sources \cite{Buckley:12} or to observe light-matter hybridisation \cite{Flatten:16_2}, has led to the use of microcavities \cite{Li:19}; specialist fabrication techniques, such as laser ablation \cite{Hunger:10} or chemical etching \cite{Trupke:05}, that can manufacture the requisite highly curved micromirrors typically produce mirrors that are not perfectly spherical \cite{Muller:10, Uphoff:15, Biedermann:10}. Secondly, in cavity optomechanics, the advantages conferred by low-mass mirrors encourage lightweight designs with limited diameter \cite{Aspelmeyer:14, Kleckner:06}. Finally, cavities with non-spherical mirrors offer useful optical capabilities, for example flexibility to tailor the optical mode \cite{Karpov:22b, Walker:21} or utilise polarisation properties \cite{Buters:16}.

As such experiments mature towards applications, it is important to calculate the required precision for transverse mirror alignment; For the spherical mirror case, there are simple methods for calculating the resonant modes under transverse mirror misalignment \cite{Hunger:10, Gao:23}, but these do not necessarily apply well to cavity mirrors with alternative shapes. This paper details extensions to the mode mixing method (Kleckner et al. \cite{Kleckner:10}), allowing for certain mirror shapes to be encoded without numerical integration, and for arbitrary mirror shapes to be transversely misaligned without further integration. These advances greatly reduce, and potentially eliminate, the computation devoted to numerical integration, allowing for the impact of transverse misalignment in cavities with deformed mirrors to be investigated thoroughly.

First, we present an intuitive geometric optics approach to predicting the modes of cavities with misaligned and non-spherical mirrors. We then overview the existing mode mixing method before detailing extensions that greatly simplify the calculations required to model particular mirror shapes, and to include transverse mirror misalignment. We then discuss geometric transformations of cavity modes that can be used to interpret calculation outputs. Finally we compare these methods to existing techniques, demonstrating good agreement with published results for Gaussian-shaped mirrors in aligned configurations while additionally permitting the easy exploration of the impact of mirror misalignment. In a further publication~\cite{Hughes:23_2}, we use the methods developed in this manuscript to examine the behaviour of cavities with spherical and Gaussian mirrors under transverse misalignment.


\section{Geometric Analysis of Mode Deformation}
\label{sec:geometry}

\begin{figure*}
\centering\includegraphics[width=0.9\textwidth]{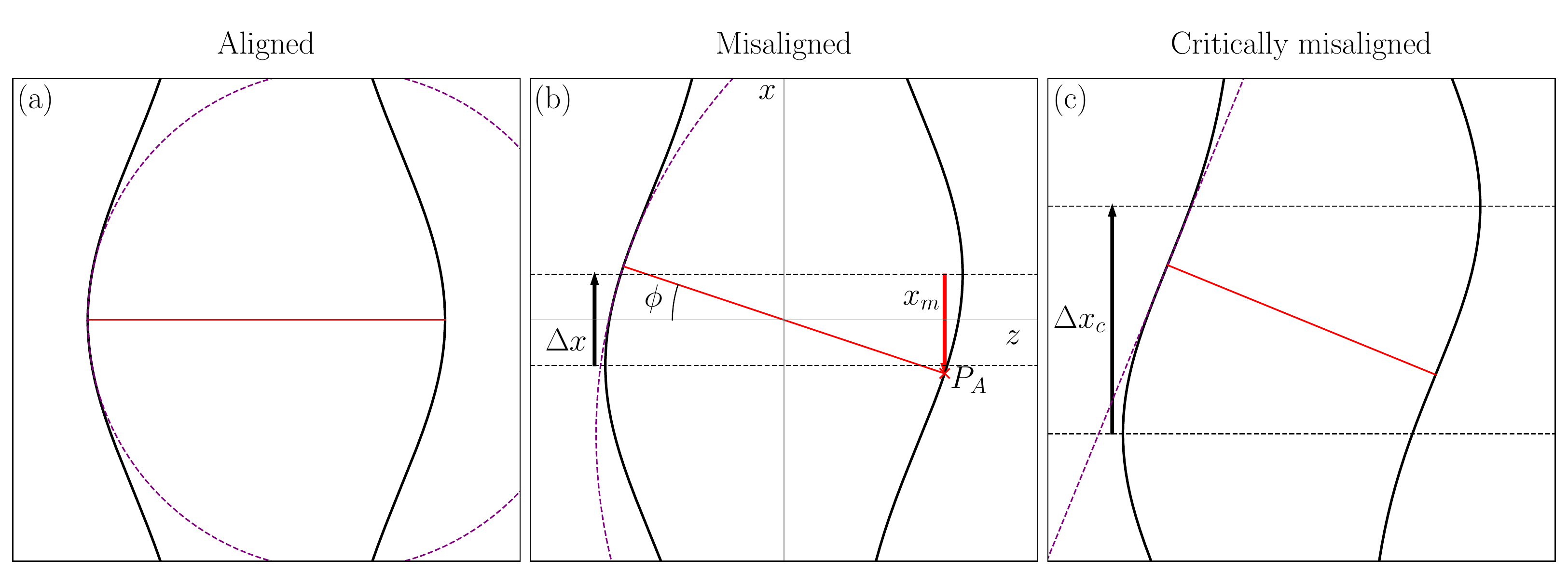}
\caption{Schematic of cavity mode geometry in misaligned optical cavities, shown for the example case of Gaussian-shaped mirrors. (a) Perfect, symmetric alignment of the two mirrors. The cavity mode axis (red) is aligned to both mirror axes. The radius of curvature of the phase fronts on the mirror matches the central curvature (purple) (b) Small transverse misalignment $\Delta x$ of the mirrors. The mode tilts at an angle $\phi$ to the cavity axis, intersecting the mirror at point $P_A$, which is at displacement $x_m$ from the centre of the mirror. The curvature at the intersection point (purple) is different from the centre of the mirror. (c) Mode instability for large mirror displacement. At sufficiently large misalignment, the cavity mode axis may tilt at such an angle that neither mirror surface is concave at the mode intersection point, and no stable mode is predicted.}
\label{fig:schematic}
\end{figure*} 

Before introducing our novel approach to mode mixing calculations in cavities with deformed mirrors and residual misalignment, we review the problem with a simple geometric optics picture that serves to highlight the physics of misaligned cavities in a more intuitive, albeit less complete, manner. In this `geometric' approach to determining the cavity modes, the propagation axis of the mode must intersect both mirrors normal to their surface, so that the mode is perfectly retroreflected. The phase curvature of the cavity mode at the intersection with each mirror is then matched to the local curvature of the mirror about the intersection point, as described in \cite{Blows:98}. This condition determines the positions and sizes of the transverse waists of the cavity mode in both transverse directions.

We consider the features predicted when applying this approach to Fabry-P\'erot cavities whose mirrors are transversely misaligned such that they are no longer coaxial. Although the method is applicable to very general mirror profiles, we will assume for simplicity that the mirror profile is a spherically symmetric depression, and we will illustrate the predicted phenomena using Gaussian shaped mirrors as a specific example, as depicted in Fig~\ref{fig:schematic}. Gaussian mirrors have a depth profile
\begin{equation}
    f_G(x,y) = D\left[1-\exp(-\frac{x^2+y^2}{w_e^2})\right],
    \label{eq: Gaussian mirror profile}
\end{equation}
where $x$ and $y$ are Cartesian coordinates transverse to the mirror axis, $D$ is the depth of the mirror, and $w_e$ the $1/e$ waist. These parameters define the central radius of curvature $R_c = w_e^2/2D$. By convention, the depth profile is zero at the centre of the depression, and positive as the concave mirror protrudes towards the centre of the cavity.

Figure \ref{fig:schematic}(a) shows the case of perfect alignment. The predicted mode lies along both (colinear) mirror axes, with the wavefront curvature at each mirror matching the centre radius of curvature $R_c$. The corresponding fundamental Gaussian mode can be calculated using standard spherical cavity theory \cite{Yariv:91}. Note that this yields a poor approximation of the fundamental mode if the mirror shape deviates significantly from spherical over the scale of the mode.

If the cavity mirrors are transversely misaligned, as shown Fig.\ \ref{fig:schematic}(b), the cavity mode axis must tilt so that it can intersect both mirrors at normal incidence. This means that the \textit{local} radius of curvature of the mirrors at the position of intersection may differ from $R_c$, producing a mode with a different waist compared to a cavity with aligned mirrors. Moreover, the local radius of curvature may differ in the two transverse directions making the cavity mode an \textit{elliptical} Gaussian beam.

To analyse these effects quantitatively, we construct a coordinate system in which the centres of the two mirrors, labelled A and B, are placed at coordinates $z_A=L/2$ and $z_B=-L/2$ respectively along the $z$ axis, where $L=z_A-z_B$ is the cavity length and the $z$ axis is the cavity axis in the aligned configuration. The misalignment direction is taken to define the $x$-axis, and thus the two mirrors are displaced by $\pm \Delta x/2$ in the $x$-direction respectively, as shown in Fig.~\ref{fig:schematic}\textbf{b)}. The point $P_A=(x_{P_A},y_{P_A},z_{P_A})$ where the cavity axis intersects mirror A can be calculated from the requirement that the cavity axis is locally orthogonal to the mirror; with $x_m=x_{P_A}-\Delta x/2$ defined as the distance of point $P_A$ from the centre of the mirror, the solution satisfies
\begin{equation}
    \frac{\Delta x}{2} = 2D\frac{x_m}{w_e^2} e^{-x_m^2/w_e^2}
    \left[\frac{L}{2}+D\left(-1+e^{-x_m^2/w_e^2}\right)\right] - x_m,
    \label{eq: mirror intersect implicit}
\end{equation}
which can be solved numerically for $x_m$ and then used to calculate the coordinates of $P_A$. 

With the mode axis determined, the properties of the cavity mode can be simply derived. The effective length of the cavity mode between the intersections with the mirror is
\begin{equation}
    \leff = 2\sqrt{x_{P_A}^2+z_{P_A}^2}.
\end{equation}
The radius of curvature of the mirror at $P_A$ in $x$ direction is
\begin{subequations}
\begin{align}
R_x & =  \frac{\left[1+f_G^{\prime 2}(x_m, 0)\right]^{3/2}}{f_G^{\prime\prime}(x_m, 0)}, \\
f_G^\prime(x_m, 0) & =  2D\frac{x_m}{w_e^2}e^{-x_m^2/w_e^2}, \\
f_G^{\prime\prime}(x_m, 0) & =  
 D\frac{2}{w_e^2}e^{-x_m^2/w_e^2}\left[1-\left(\frac{\sqrt{2}x}{w_e}\right)^2\right]
\end{align}
\label{eq: radius of curvature parallel to displacement}
\end{subequations}
where $f_G^\prime(x,y)$ and $f_G^{\prime\prime}(x,y)$ are first and second derivatives of the mirror profile $f_G(x,y)$ (Eq.~\ref{eq: Gaussian mirror profile}) with respect to $x$. The radius of curvature in the $y$ direction is
\begin{subequations}
\begin{align}
R_y & =  R_c e^{x_m^2/w^2} \cos\phi + x_m \sin\phi, \\
\sin\phi & =  \frac{x_{P_A}}{\sqrt{x_{P_A}^2+z_{P_A}^2}},
\end{align}
\label{eq: radius of curvature perpendicular to displacement}
\end{subequations}
where $\phi$ is the angle of the cavity mode axis with respect to the $z$ axis. The central waists are
\begin{equation}
    w_{0,v} = \sqrt{\frac{\lambda \leff}{2\pi}}\left(\frac{2R_v}{\leff}-1\right),
\end{equation}
where $v\in \{x,y\}$ specifies the transverse coordinate\footnote{The principal axes of the mode will be in the $x$ and $y$ directions because the transverse misalignment is $x$-directed.}.

For large mirror misalignments, the mode axis may intersect the mirror sufficiently far from the central depression that the local profile is not concave, as shown Fig.~\ref{fig:schematic}(c). In this case the cavity is not able to stably confine a mode. For Gaussian mirrors, this occurs for misalignments $\Delta x$ exceeding $\Delta x_c$ at which $x_m=\pm w_e/\sqrt{2}$.

\begin{figure}
\centering\includegraphics[width=0.7\columnwidth]{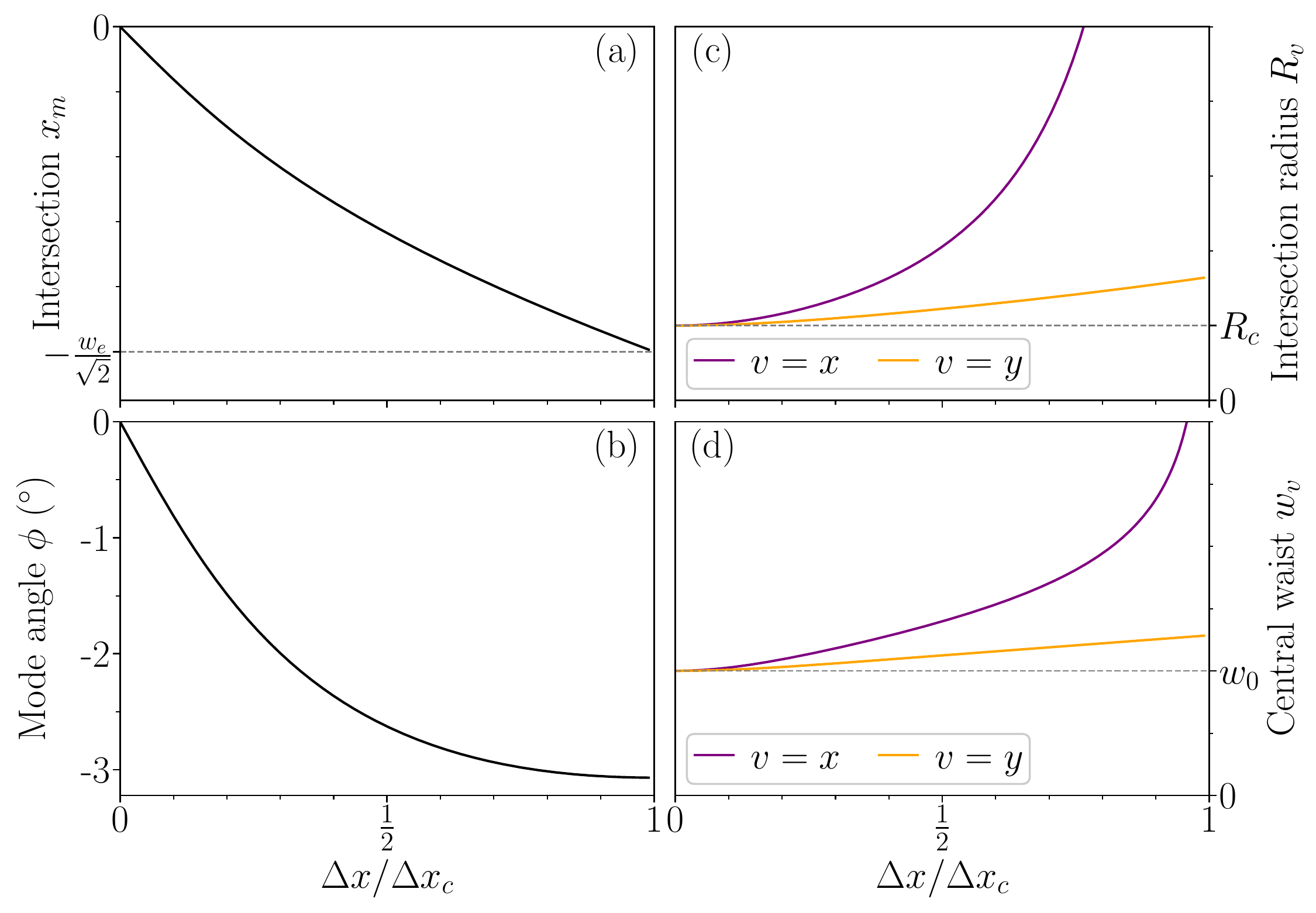}
\caption{Case study of the predicted mode in cavities with Gaussian-shaped mirrors under transverse misalignment, generated for a cavity with $L=500~\upmu\mathrm{m}$, $R_c=400~\upmu\mathrm{m}$, $w_e=50~\upmu\mathrm{m}$, with interrogation wavelength \SI{866}{\nano \metre}. (a) Intersection coordinate relative to the centre of the mirror. (b) Tilt angle $\phi$, (c) Local radii of curvature at the mode intersection $R_{v}$ for $v \in \{x, y\}$, and (d) predicted central waists in the $x$ and $y$ directions.}
\label{fig:geometry}
\end{figure}

A numerical case study applying this procedure to a cavity with Gaussian-shaped mirrors is presented in Fig.~\ref{fig:geometry}. This shows that, as the mirrors are misaligned, the mode angle and the position of intersection on the mirror deviate increasingly from their aligned values. The off-axis intersection means that the local radius of curvature at the intersection points increases in both $x$ and $y$ directions. However, the change is much larger in $x$ direction. At the critical misalignment $\Delta x_c$ (\SI{44.0}{\micro\metre} for the parameters of Fig.~\ref{fig:geometry}), the mode intersection point is sufficiently far from the centre of the mirror that the local mirror surface is not concave. This means that the cavity is unstable, and one would expect to observe a severe drop in finesse.

This `geometric' analysis of the fundamental mode limits itself to cavity modes with quadratic wavefront curvature, and therefore does not take account of the mirror shape beyond its local gradient and radius of curvature. Though the mirror surface can always be approximated as parabolic close enough to the intersection point, the geometric analysis becomes unsuitable when the mode is sufficiently wide on the mirror that higher-order components of the profile become significant. To calculate cavity modes for cases where the mirror profile is not perfectly parabolic about the mode intersection points, we must use a framework with the flexibility to model cavity modes with more general wavefront curvature profiles.


\section{Extended Mode Mixing Method}
\label{sec: Theory mirror transformation}
\subsection{Mode Mixing Introduction}
\label{subsec: mode mixing introduction}
The mode mixing method \cite{Kleckner:10} finds the stable modes of cavities with deformed mirrors by expressing propagating fields as linear superpositions of Gaussian modes. This method has been applied to microcavities with non-spherical mirrors, finding sporadic, severe drops in cavity finesse at particular cavity lengths due to resonant mixing of the basis modes \cite{Benedikter:15, Benedikter:19}. Alternatively, mode mixing can be harnessed to increase coupling of cavity fields to single emitters \cite{Podoliak:17, Karpov:22, Karpov:22a}, introduce coupling between optical resonators \cite{Flatten:16} or tailor cavity modes to have desired properties \cite{Karpov:22b}. Standard mode mixing theory is introduced in this section, before extensions to facilitate the calculations, particularly in the context of misaligned cavities, are presented.

In principle, a propagating electric field satisfies Maxwell's equations. Typically, these equations are simplified by employing the paraxial approximation, which assumes that the propagating field is beam-like and directed at small angles to the nominal $z$ axis. Under these assumptions (see \cite{Barre:17}, with which the notation presented is consistent), the electric field can be described via a scalar function $u^{\pm}$ through
\begin{equation}
    \bm{E}(x,y,z,t) = \bm{\epsilon} u^{\pm}(x,y,z)\exp(\mp ikz)\exp(i\omega t),
\label{eq: vector mode to scalar}
\end{equation}
where $\omega$ is the angular frequency, $k=\omega/c$ the wavevector, $\bm{\epsilon}$ the constant linear polarisation of the field, which must lie in a plane perpendicular to the $z$-axis, and $\pm$ denotes propagation towards positive or negative $z$ respectively. The function $u^{\pm}$ satisfies the paraxial wave equation
\begin{equation}
    \pdv{z}u^{\pm}(x,y,z) = \mp\frac{i}{2k}\left(\pdv[2]{x} + \pdv[2]{y}\right)u^{\pm}(x,y,z).
    \label{eq: paraxial equation}
\end{equation}

\begin{figure}
\centering\includegraphics[width=0.6\columnwidth]{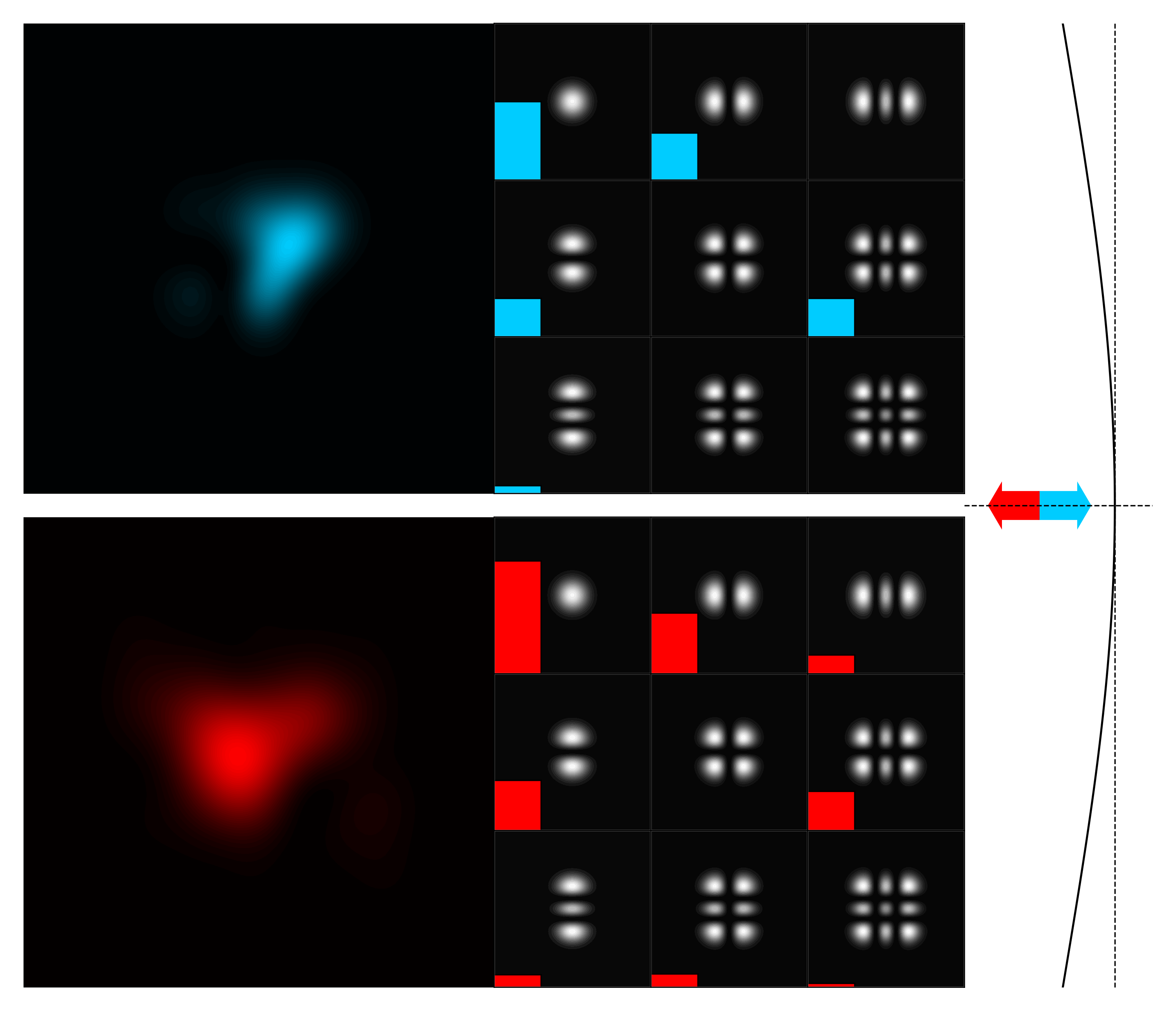}
\caption[Schematic of mode mixing method]{Diagram of the mode mixing introduced by reflection from a mirror. The incoming beam (blue, top) is expressed as a linear superposition of incoming basis states (here the basis states for $n_x$ and $n_y$ up to 2 are shown). The amplitudes of the coefficients are depicted by the height of the bars adjacent to the corresponding basis states. The mirror (solid black concave profile), shown as having a Gaussian profile, imprints a phase front on the input beam and reverses its direction. The output beam (red, bottom) is expressed as a linear superposition of basis states in the outgoing basis. The intensity patterns of the input and output beams are plotted in the transverse plane at which both bases have their central waist.}
\label{fig: mode mixing explanation}
\end{figure}

In the mode mixing formalism, an electromagnetic field propagating along the $z$ axis according to Eq~(\ref{eq: paraxial equation}) is expressed as a linear superposition of modes $u^{\pm}_s(x,y,z)$, which themselves satisfy the paraxial equation, where $s$ is an index over all the modes in the basis. An optical element is encoded as a matrix whose elements are scattering amplitudes from ingoing modes in the ingoing basis to outgoing modes in the outgoing basis. In the case of a concave mirror illuminated at normal incidence, the input and output basis states counterpropagate and the mirror profile imprints a differential phase across the wavefront due to the variation in propagation distance to and from the mirror. The components of a mirror matrix $A$ ($B$) at positive (negative) $z$ coordinate may be written
\begin{subequations}
\begin{align}
    A_{s,t} = & \int_{S_A} u^{-*}_s(x,y,z_A) \exp\left(2ik f_A(x,y)\right) u^{+}_t(x,y,z_A) dS, \\
    B_{s,t} = & \int_{S_B} u^{+*}_s(x,y,z_B) \exp\left(2ik f_B(x,y)\right) u^{-}_t(x,y,z_B) dS,
\end{align}
    \label{eq: traditional mode mixing action of mirror}
\end{subequations}
where $k$ is the wavevector of the light, $z_A$ ($z_B$) is the axial coordinate of the centre of the depression of mirror A (B), $f_A$ ($f_B$) is the surface profile of mirror A (B) (with the convention that a positive profile points towards the cavity centre for both mirrors) and $S_A$ ($S_B$) is the surface region of mirror A (B). The surface integrals are each performed in a single transverse plane at the axial coordinate for which $f_A$ is zero. A schematic diagram illustrating how a mirror transfers amplitude from the input basis to the output basis is shown in Fig.~\ref{fig: mode mixing explanation}. Cavity eigenmodes are specific linear superpositions of basis states that are preserved after one round trip of a cavity.

In this manuscript, the basis states used to express the cavity function $u^{\pm}(x,y,z)$  are the Hermite-Gauss modes
\begin{equation}
u^{(\pm)}_{n_x,n_y}\left(x,y,z\right)  =  a(z) H_{n_x}\left(\frac{\sqrt{2}x}{w(z)}\right)H_{n_y}  \left(\frac{\sqrt{2}y}{w(z)}\right)
\exp\left[-\frac{x^2+y^2}{w(z)^2}\right] \exp\left[\mp ik\frac{x^2+y^2}{2R_u(z)}\right]\exp\left[\pm i(n_x+n_y+1)\Psi_G\right]
\label{eq: Hermite Gauss mode}
\end{equation}
where
\begin{equation}
    \begin{aligned}
        a(z)  =  \frac{1}{w(z)} \sqrt{\frac{2}{\pi}\frac{1}{2^{n_x+n_y} n_x!n_y!}},  \quad
w(z)  = & w_0 \sqrt{1+\left(\frac{z}{z_0}\right)^2}, \\
z_0 = \frac{\pi w_0^2}{\lambda}, \quad
R_u(z)  = z\left(1+\left(\frac{z_0}{z}\right)^2\right), \quad
& \Psi_G(z)  =  \arctan\left(\frac{z}{z_0}\right),
    \end{aligned}
\end{equation}
where the wavelength $\lambda=2\pi/k$, $H_i$ are the Hermite polynomials with $n_x, n_y \in \mathbb{N}$ the $x$ and $y$ transverse indices, and $z_0$ is the Rayleigh range of the beam. This basis is complete and orthonormal for each transverse plane separately. A cavity function expressed as a linear superposition of these basis modes retains its mode coefficients during propagation, as the propagation of the field is encoded in the $z$-dependence of the basis functions themselves. 

The round-trip matrix can be calculated from the two mirror matrices, accounting for the round trip phase accumulated during propagation:
\begin{equation}
    M = BAe^{-2ikL}.
\end{equation}
A mode $\ket{\Psi_i}$ supported by the cavity  is an eigenmode of the round-trip matrix $M$, and has corresponding eigenvalue $\gamma_i$ from the eigenmode equation
\begin{equation}
    M\ket{\Psi_i} = \gamma_i\ket{\Psi_i}.
\end{equation}
The complex $\gamma_i$ has both phase and amplitude. The complex phase is the round-trip phase (modulo $2\pi$) accrued by $\ket{\Psi_i}$, which is zero on resonance. For typical applications where the length can be tuned freely to match a given resonance, the amplitude is more pertinent  as it leads directly to the round-trip loss $\loss{RT} = 1-\abs{\gamma_i}^2$.

The eigenmodes of cavities with deformed mirrors can be determined by calculating elements of mirror matrices $A$ and $B$ through integration of Eq.~(\ref{eq: traditional mode mixing action of mirror}). A sensible approach to calculating the eigenmodes of cavities with transverse misalignment would therefore appear to calculate the geometrically-expected mode as a function of misalignment using the theory of Sec.~\ref{sec:geometry}, and use this mode to define the basis of the mode mixing calculation. Using this approach, the basis of the mode-mixing calculation is always chosen to suit the geometric model, and therefore it should be easier to faithfully capture the cavity eigenmodes with a relatively limited basis size.

However, performing calculations this way uses a different basis for every misalignment and cavity length. Therefore, all of the matrix elements are calculated for each cavity configuration separately. An alternative approach, discussed for the remainder of this section, uses matrix operations to misalign the mirrors without changing their calculation basis, thus removing the need to explicitly encode the mirror profiles for every misalignment.

\subsection{Replacing Coordinates with Operators}
The long-appreciated similarities between the Hermite-Gauss modes and simple harmonic oscillator wavefunctions \cite{Stoler:81, Nienhuis:93} inspire the writing of transverse coordinates $x$ ($y$) and transverse derivatives $\partial/\partial x$ ($\partial/\partial y$) in terms of the ladder operators $a_x$ ($a_y$), where $a_x$ ($a_y$) reduces the $n_x$ ($n_y$) index of the Hermite Gauss mode by 1. Such operator methods have already been used to determine the eigenmodes of optical cavities under particular circumstances \cite{Habraken:07, Jaffe:21, vanExeter:22}. According to the conventions of the present analysis, the operators for $x$ and $\partial/\partial x$ in a given transverse plane are

\begin{subequations}
\begin{align}
x^{(\pm)}(z) & =  U_G^{(\pm)}(z)^{\dag}\frac{1}{2}w(z)(a_x+a^{\dag}_x)U_G^{(\pm)}(z), \\
\frac{\partial}{\partial x} & =  \frac{1}{w_0}(a_x-a^{\dag}_x), \\ 
(U_G)^{(\pm)}(z)_{n_x',n_y',n_x,n_y} & =  e^{\pm i\Psi_G(z)(n_x+n_y+1)} \delta_{n_x',n_x}\delta_{n_y',n_y}, 
\end{align}
\label{eq: position operator}
\end{subequations}
where $n_x$ and $n_y$ ($n_x'$ and $n_y'$) are the $x$ and $y$ indices of the input (output) modes of the matrix respectively. While the $\partial/\partial x$ operator does not depend on propagation direction and is constant across all transverse planes, the matrix elements of $x$ depend upon the $z$ coordinate and the propagation direction. The equivalent relations hold for $y$ and $\partial/\partial y$, with $a_y$ ($a^{\dag}_y$) replacing $a_x$ ($a^{\dag}_x$). The derivations are detailed in App.~\ref{app: derivation_of_HG_operators}.

A mirror imprints a phase front onto and reflects the ingoing mode (as expressed in Eq.~(\ref{eq: traditional mode mixing action of mirror})). To construct mirror matrices in an operator-based approach, it is conceptually simpler to consider this process sequentially (taking mirror A as the example case): First, the phase front $\exp(2ikf_A)$ is imprinted on the input basis, where the phase is no longer a complex function of coordinates $x$ and $y$, but an \emph{operator} acting on the input basis as a result of its composition in the coordinate operators $x$ and $y$. Secondly, the reflected field, thus far expressed through coefficients in the input basis, is transferred to coefficients in the output basis through operator 

\begin{equation}
U^{+\rightarrow -} = (U_G^{(+)})^2 \exp\left(-2ik\frac{(x^{(+)})^2 + (y^{(+)})^2}{2R_u(z_A)}\right),
\end{equation}
for mirror A and

\begin{equation}
U^{-\rightarrow +} = (U_G^{(-)})^2 \exp\left(-2ik\frac{(x^{(-)})^2 + (y^{(-)})^2}{2R_u(z_B)}\right),
\end{equation}
for mirror B, where $R_u(z_A)$ and $R_u(z_B)$ depend upon the chosen basis. This basis is most conveniently chosen so that the wavefront radius of curvature $R_u(z_A)$ ($R_u(z_B)$) matches the radius of curvature $R_A$ ($R_B$) of the quadratic component of the profile of mirror A (B). This choice uniquely specifies the basis, and is assumed for the remainder of the text. The mirror matrix $A$ can then be expressed

\begin{equation}
A = (U_G^{(+)})^2 \exp\left(-2ik\Delta_A^{(+)}\right),
\label{eq: mirror matrix profile and reflection}
\end{equation}
where $\Delta_A = f_A - (x^2 + y^2)/2R_A$ is the deviation of the profile of mirror A from the ideal parabolic surface\footnote{In the paraxial approximation, the mathematically ideal mirror profile is parabolic. Outside this approximation, a spherical mirror is is often a better match for the phase fronts \cite{Laabs:99}.}. If $\Delta_A$ ($\Delta_B$) can be evaluated as a matrix without taking integrals, the mirror matrix $A$ ($B$) can also be obtained without integrals, as discussed later in Sec.~\ref{subsec: Taking exponent of surface profile}.

\subsection{Calculating Polynomial Mirror Surface Profiles} 
For the case where $\Delta$ can be written as a power series in $x$ and $y$, it is only necessary to calculate matrices of the various powers of $x$ and $y$ and sum each polynomial term with the appropriate coefficient. For the case of a parabolic distortion, the mirrors remain parabolic but with an adjusted radius of curvature, and therefore the cavity eigenmodes should match standard results. We have used this to test and validate our approach. 

\subsection{Calculating the Gaussian Surface Profile}
\label{subsec: calculate Gaussian surface profile}
The Gaussian surface profile can also be expressed in the Hermite-Gauss basis without taking integrals, but this requires a different approach, inspired by the appendix of \cite{Varro:22} and detailed in App.~\ref{app: finding_gaussian_profile_matrix}. The matrix elements of a unit Gaussian profile with $1/e$ waist $w_e$ in a one-dimensional Hermite-Gauss basis at axial coordinate $z$ can be written

\begin{equation}
\exp\left[-\frac{x^2}{w_e^2}\right]_{m',m}^{(\pm)}(z) =  U_G^{(\pm)}(z)^{\dag} \left(1-\chi\right)^{-(\frac{m'+m+1}{2})}  \left(\frac{\chi}{2}\right)^{\frac{m'-m}{2}}\sqrt{m'!m!}\sum_{k=0}^{[\frac{m}{2}]} \frac{\left(\frac{\chi^2}{4}\right)^k}{\left(\frac{m'-m}{2}+k\right)!k!\left(m-2k\right)!} U_G^{(\pm)}(z),
\label{eq: Gaussian surface operator}
\end{equation}
with 
\begin{equation}
    \chi =  -\frac{1}{2}\frac{w\left(z\right)^2}{w_e^2}, 
\end{equation}
where $m$ ($m'$) is the index of the ingoing (outgoing) mode, $(m'-m)/2$ is an integer and $m' \geq m$. If $(m'-m)/2$ is not an integer, the matrix element is zero. If $m > m'$, the symmetry $\exp[-x^2/w_e^2]_{m',m} = \exp[-x^2/w_e^2]_{m,m'}$ should be used. The matrix $U_G$ accounts for the Gouy phases of the basis states, as originally defined in Eq.~(\ref{eq: position operator}). The two-dimensional profile is obtained from the one-dimensional matrices by a simple tensor product.

The deviation matrix $\Delta$ of a Gaussian with depth $D$ from the ideal parabolic surface is obtained from the matrix of the unit profile through

    \begin{equation}
       \Delta^{(\pm)} = D\left(1-\exp\left[-\frac{(x^{(\pm)})^2+(y^{(\pm)})^2}{w_e^2}\right]\right)-\frac{(x^{(\pm)})^2+(y^{(\pm)})^2}{2R}, \label{eq: Gaussian delta} 
    \end{equation}
with
\begin{equation}
D = \frac{w_e^2}{2R} \label{eq: Gaussian D},
\end{equation}
where $R$ is the radius of curvature at the centre of the Gaussian. The use of a single $R$ in Eq.~\ref{eq: Gaussian delta} and Eq.~\ref{eq: Gaussian D} imposes that the wavefront radius of curvature of the basis states matches the mirror radius of curvature in the central depression.

\subsection{Taking the Exponent of the Surface Profile}
\label{subsec: Taking exponent of surface profile}
Once the surface profile deviation $\Delta$ is expressed as a matrix, the surface profile phase matrix $\exp\left(-2ik\Delta\right)$, which constitutes the non-trivial component of the mirror matrix (Eq.~\ref{eq: mirror matrix profile and reflection}), can be calculated. It is tempting to calculate $\exp\left(-2ik\Delta\right)$ through matrix exponentiation of $-2ik\Delta$, but this method cannot model losses; as $\Delta$ is a Hermitian matrix, the matrix exponent is unitary, and therefore every eigenvalue of a mirror matrix obtained through matrix exponentiation has unit modulus, meaning that the mirror is lossless. No matter how large a basis is chosen, $\Delta$ never models processes representing transfer from inside to outside the basis, and therefore no mechanism exists for power to leave the cavity.

To take the exponential in a way that can model losses, A procedure is used which is conceptually similar to the non-Hermitian Hamiltonian approach to simulating quantum systems that is commonly used in cavity quantum electrodynamics \cite{Kuhn:15}. The matrix $\Delta$ is first evaluated in a basis larger than the intended simulation basis, before being truncated to the size of the simulation basis according to specific rules: Each element of $\Delta$ represents a transfer from an input state to an output state. If the input state lies within the simulation basis, but the output state is outside, that element encodes loss. Therefore, for each input state, the sum over all the magnitudes of transfers to states outside the basis is calculated, evaluating the amplitude leakage from the input basis state to outside the simulation basis. This summed rate is then added as a negative imaginary number onto the diagonal element of the input state. When the matrix exponential is then taken, this diagonal imaginary component causes loss rather than amplitude transfer.

Expressed mathematically, for a larger basis containing $n_x$ and $n_y$ up to maximum values of $n_x^{\mathrm{H}}$ and $n_y^{\mathrm{H}}$ respectively, and the smaller simulation basis up to maximum values of $n_x^{\mathrm{NH}}$ and $n_y^{\mathrm{NH}}$ respectively, components of the non-Hermitian $\Delta$ matrix are written
\begin{subequations}
\begin{align}
\Delta^{(\pm)}_{n_x',n_y',n_x,n_y} & = \Delta^{\mathrm{H}(\pm)}_{n_x',n_y',n_x,n_y},\, \, \, \, \, \, \,  n_x',n_x\leq n_x^{\mathrm{NH}},\, \, n_y',n_y\leq n_y^{\mathrm{NH}}, \, \, \delta_{n_x',n_x}\delta_{n_y',n_y} = 0, \\
\Delta^{(\pm)}_{n_x,n_y,n_x,n_y} & = \Delta^{\mathrm{H}(\pm)}_{n_x,n_y,n_x,n_y} + i \sum_{n_x' = (n_x^{\mathrm{NH}}+1)}^{n_x^{\mathrm{H}}}\;\sum_{n_y' = (n_y^{\mathrm{NH}}+1)}^{n_y^{\mathrm{H}}} |\Delta^{\mathrm{H}(\pm)}_{n_x',n_y',n_x,n_y}|,
\end{align}
\end{subequations}
where $\Delta^{\mathrm{H}(\pm)}$ is the Hermitian surface profile deviation matrix evaluated on the larger basis. The matrix exponential of the non-Hermitian $\Delta$ is then taken to find surface profile phase matrix $\exp\left(-2ik\Delta\right)$.

While this process is not mathematically identical to finding the true matrix $\exp\left(-2ik\Delta\right)$, in practice, this procedure produces almost identical loss results to numerical integration for most cavity configurations, as shown later in Sec.~\ref{sec: demonstration}.

\subsection{Translating the Mirror}
\label{subsec: Translating mirror}
With the surface profile phase matrix calculated, it is possible to evaluate both mirror matrices and thus obtain the eigenmodes for a cavity. To investigate the impact of transverse misalignment between the mirrors, the mirror matrices could be calculated for every misalignment separately. An alternative, discussed in this section, is to evaluate the mirror matrix in one transverse position (most conveniently the aligned configuration where any symmetries of the mirror profile can be exploited) and use translation operators to model transverse misalignment without calculating any further mirror matrix elements directly. 

\begin{figure}
\centering\includegraphics[width=0.6\columnwidth]{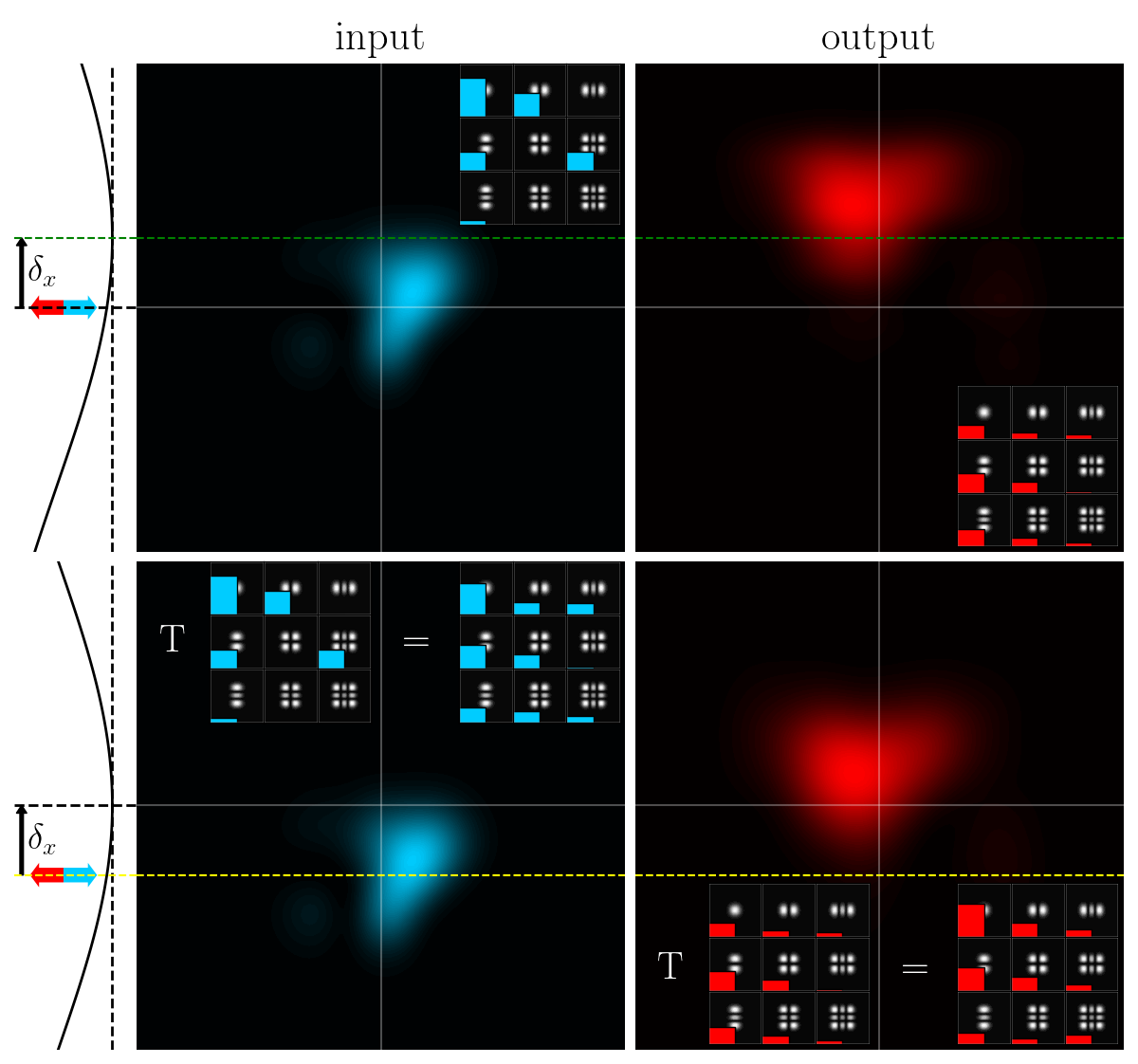}
\caption[Schematic of transverse translations in mode mixing method]{Explanation of the method for obtaining the mirror matrix of a mirror translated in the $x$ direction by $\delta_x$. Top: The input mode (blue, left) hits a translated mirror (black concave surface) and scatters to the output mode (red, right). Both input and output modes are expressed as a linear superposition of basis states (here shown up to $n_x$ and $n_y$ of 2). Bottom: The same physical process described in a coordinate system centred on the mirror, such that the mirror is nominally in the aligned configuration. Due to the shift in origin, the input and output fields have been translated using translation operator $T$. The physical equivalence of the two scenarios means that the action of the translated mirror can be derived from the untranslated mirror through suitable transformation of the input and output bases.}
\label{fig: mirror translation explanation}
\end{figure}

As depicted in Fig.~\ref{fig: mirror translation explanation} the action on a given input field of a mirror translated by $\delta_x$ in the $x$ direction is equivalent to the action of the untranslated mirror on the same input field displaced by $-\delta_x$, because these two cases describe the same physical situation for different choices of origin. This equivalence means that the matrix of the translated mirror can be calculated by taking the matrix of the untranslated mirror and translating the input and output bases in the compensating direction.

The one-dimensional operator that translates the input and output bases is
\begin{equation}
    T\left(\delta\right) = \exp\left(\delta \frac{\partial}{\partial x}\right),
\end{equation}
with elements
\begin{subequations}
\begin{align}
T\left(\delta\right)_{m',m}  = & \sqrt{\frac{m!}{m'!}} \alpha^{m'-m} e^{-\frac{\alpha^2}{2}}L_n^{m'-m}\left(\alpha^2\right) \, , \, m' \geq m, \\
T\left(\delta\right)_{m',m}  = &\sqrt{\frac{m'!}{m!}} \left(-\alpha^{m-m'}\right) e^{-\frac{\alpha^2}{2}}L_m^{m-m'}\left(\alpha^2\right) \, , \, m > m', 
\end{align}
\end{subequations}
where 
\begin{equation}
    \alpha = \frac{\delta}{w_0},
\end{equation}
where $m$ ($m'$) is the input (output) index in the one-dimensional basis and $\delta$ the translation effected by the operator. This operator is identical to the displacement operator of the simple harmonic oscillator \cite{Cahill:69}, owing to the close similarity between the simple-harmonic and Hermite-Gauss bases. As the translation operator has the same elements in the input and output bases, translating a mirror with matrix $C$ by $\delta_x$ can be achieved through
\begin{equation}
C \rightarrow T_x^{\dag}\left(-\delta_x\right) C T_x\left(- \delta_x\right),
\end{equation}
where $T_x$ is formed from the tensor product of the one-dimensional translation in the $x$ direction and the identity in the $y$ direction. If scanning the misalignment of the mirrors, the translation matrix need only be calculated for a single increment, and then successively applied to generate all of the mirror matrices. In this way, the mirror profile and translation step matrices both need only be calculated once.

\subsection{Mode transformations}
\label{subsec: mode transformations}

In addition to the $x$-translation operator $T_x$ discussed in the previous section, further transformation operators can be specified. Here, we present transformation operators to change the central waist of a mode, and to change its propagation angle. In the context of the current work, these operators are used not to calculate the cavity eigenmodes, but to evaluate geometric properties of these eigenmodes, as will be discussed in Sec.~\ref{sec: demonstration}. 

\subsubsection{Changing the Mode Waist}
\begin{figure*}
\centering\includegraphics[width=0.8\textwidth]{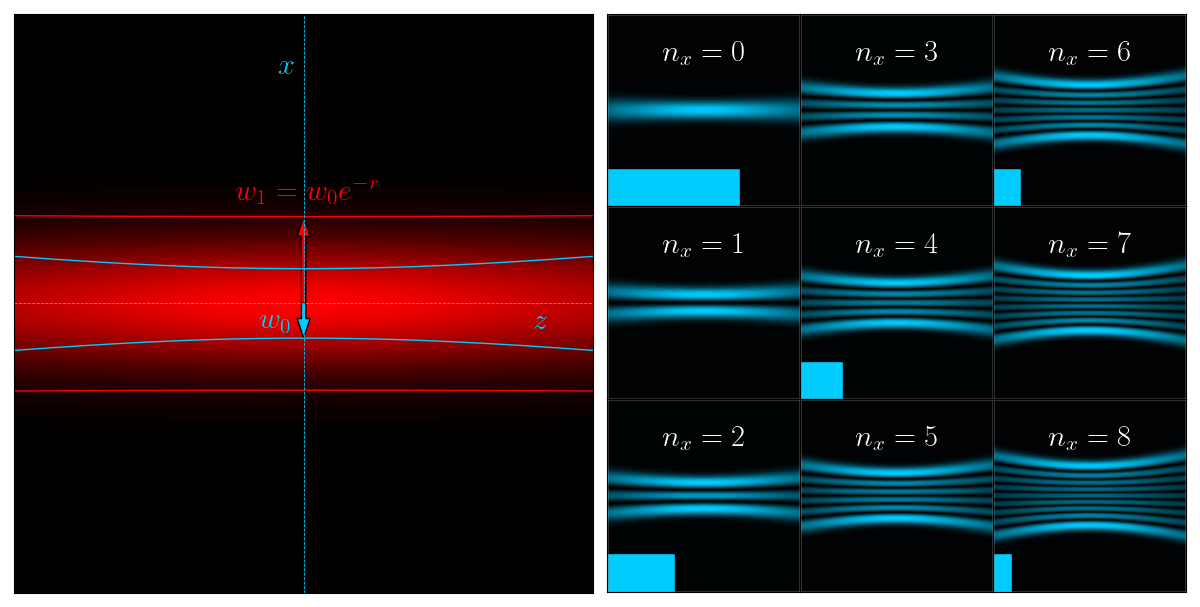}
\caption{Illustration of changing the waist of a mode through matrix methods, showing the modes in the $xz$ plane. Left: A mode with waist $w_1$ in the $x$ direction (red) is expressed in the basis of modes with waist $w_0$ in the $x$-direction (blue). Solid lines indicate the $1/e$ waist of the beam and the basis in red and blue respectively. Right: The amplitude of the coefficients of the expanded (red) mode in the original (blue) basis, where the label indicates the $n_x$ index of the mode, the image the intensity profile in the $xz$ plane and the bar the amplitude of the mode in the superposition. The rescaled fundamental mode is a linear superposition of even orders in the original basis. Note that, while only the amplitude of the coefficients of the basis modes is plotted, the phase of these coefficients is crucial.}
\label{fig: waist rescaling}
\end{figure*}

To calculate the coefficients of a mode with a different centre waist, we use the property that the Hermite-Gauss modes have the same functional form as the simple harmonic oscillator wavefunctions at the axial centre of the mode ($z=0$). Therefore, the operator that changes the central waist of the mode is the same as the operator that rescales the coordinate operators of the simple harmonic oscillator, namely the standard squeeze operators. The operator that changes the waist in the $x$-direction from $w_0$ to $w_1$ is
\begin{subequations}
\begin{align}
S_x \left(\frac{w_1}{w_0}\right) & = \exp\left[-\frac{1}{2}r\left(a_x^2 - (a_x^\dag)^2\right)\right], \\
r & =  -\log\left[\frac{w_1}{w_0}\right].
\end{align}
\end{subequations}
The use of this operator to expand the waist of a fundamental mode is depicted in Fig.~\ref{fig: waist rescaling}. The same form of operator applies in the $y$-direction for creation (annihilation) operator $a^{\dag}_y$ ($a_y$).

\subsubsection{Changing the Mode Angle}
\label{subsec: changing mode angle}
Finding the transformation operator to rotate the direction of propagation of the field is considerably more involved. This is because rotating an optical field $\bm{E}(x,y,z,t)$ is not equivalent to rotating all of the basis states $\{u_{nm}(x,y,z)\}$ due to two main complications. Firstly, as the mode envelope is rotated, the implicit axial phase $\exp(\mp ikz)$ must rotate with it. This `hidden' component will turn out to be the quantitatively dominant component of the rotation matrix. Secondly, while the optical field is a vector quantity, mode mixing is a scalar theory, with the polarisation $\bm{\epsilon}$ factoring out. The rotation operator in the mode mixing formalism rotates only the scalar field, whereas in a vector theory the rotation operator would also rotate the direction of the vector field.

With those complications noted, the operator to rotate the propagation direction can be derived. We consider an optical field $\bm{E}(x,y,z,t)$, which is a function of coordinates $x$, $y$ and $z$. Next we define a new Cartesian coordinate system in which the axes have been rotated about the $y$-axis to yield
\begin{equation}
    x' = x\cos(\phi_x) + z \sin(\phi_x), \quad y'=y, \quad z' = z\cos(\phi_x) - x \sin(\phi_x).
\end{equation}

\begin{figure*}
\centering\includegraphics[width=0.75\textwidth]{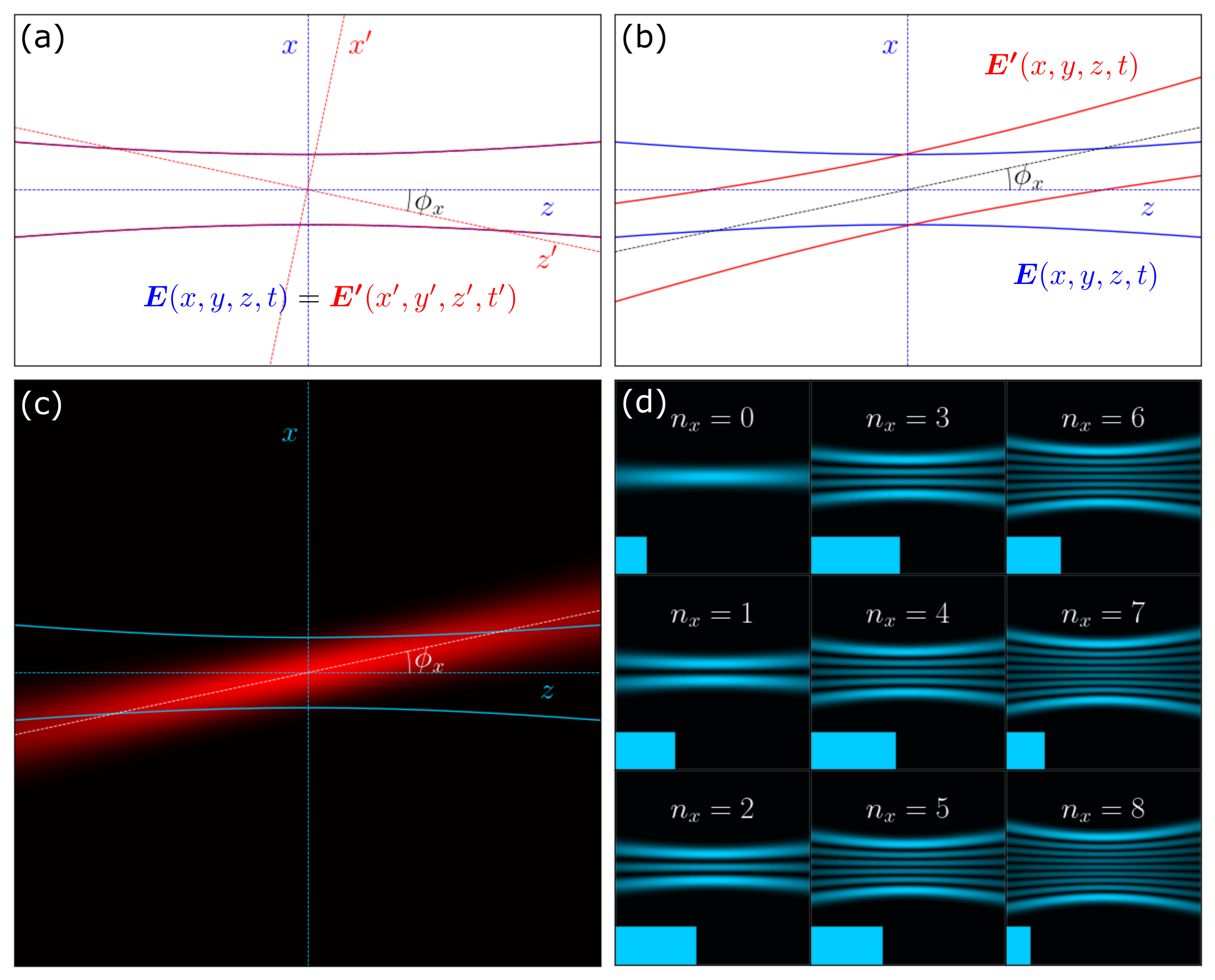}
\caption{The geometry used to derive the $xz$ rotation operator and an example of its application to rotate the propagation direction of a fundamental mode. a) A single field is described by two functions: $\bm{E}(x,y,z,t)$ in coordinates $x$, $y$, $z$ and $t$, or $\bm{E'}(x',y',z',t)$ in coordinates $x'$, $y'$, $z'$ and $t$, where the primed coordinate system has been rotated by $-\phi_x$ in the $xz$ plane. b) If the coordinate arguments of both functions are the same, $\bm{E'}(x,y,z,t)$ describes the same beam as $\bm{E}(x,y,z,t)$, but with the application of a rotation by $\phi_x$ in the $xz$ plane. c) A fundamental mode with propagation angle $\phi_x$ (red) is expressed in the basis of modes propagating along the $z$ axis with $\phi_x = 0$ (blue, with solid lines indicating the $1/e$ waist). d) The amplitude of the coefficients of the expanded (red) mode in the original (blue) basis for $n_x\in [0,8]$, where the image depicts the intensity profile of the basis states in the $xz$ plane and the bar the amplitude of the basis state in the superposition.}
\label{fig: mode rotation}
\end{figure*}

The same optical field can be expressed in the new coordinate system through the function $\bm{E'}(x',y',z',t)$. The function $\bm{E'}$ encodes the same field as $\bm{E}$, but, in its basis, the propagation direction is rotated towards the $x'$ axis in the $x'-z'$ plane. Therefore the transformation that takes the function $\bm{E}$ to $\bm{E'}$ is the operator for the propagation direction rotation, provided the coordinate arguments to both functions are the same. The coordinate systems used to derive the propagation direction-rotation operator, and the application of this operator to rotate the propagation direction of a mode, are depicted in Fig.~\ref{fig: mode rotation}.

The equivalence of $\bm{E}$ and $\bm{E'}$ in real space means that
\begin{equation}
    \bm{E'}(x',y',z',t) = \bm{E}(x,y,z,t).
\end{equation}
Now, we assume that the rotation angle is small, and thus denoted $\delta \phi_x$. As, in the conventions of this manuscript, the mode coefficients are not functions of the axial coordinate, any axial coordinate $z'$ could be chosen, but for algebraic convenience we choose the $z'=0$ plane. A first order approximation yields
\begin{equation}
\begin{aligned}
    x=x', \quad y&=y', \quad z=x'\delta\phi_x, \\
    \bm{E'}(x',y',z'=0,t) &= \bm{E}(x=x',y=y',z=x'\delta\phi_x,t).
\end{aligned}
\end{equation}
Remembering that the electric field $\bm{E}$ is described by mode function $u^{(\pm)}$ through Eq.~(\ref{eq: vector mode to scalar}) (and equivalently for $\bm{E'}$ and $u'^{(\pm)}$)
\begin{equation}
    u'^{(\pm)}(x',y',z'=0) = u^{(\pm)}(x=x', y=y', z=x'\delta\phi_x)\exp(\mp ik(z=x'\delta\phi_x)).
\end{equation}
Using the first order expansions in $\delta\phi_x$ we obtain 
\begin{equation}
    u'^{(\pm)}(x,y,0) = \left[1+x\delta\phi_x \left(\mp ik + \frac{\partial}{\partial z}\right)\right]u^{(\pm)}(x,y,0),
\end{equation}
where $x'=x$ and $y'=y$ have been used to unify the function arguments. This therefore expresses the transformation of the basis functions associated with infinitesimal rotation of the electric field.

For finite rotations, the infinitesimal operator can be applied successively, and existing results can make the final form more useful. Firstly, the $x$-operator in the $z=0$ plane is $x|_{z=0}=(kw_0/2)(a_x+a_x^{\dag})$ (see Eq.~(\ref{eq: position operator})) Secondly, the basis functions satisfy the paraxial equation (Eq.~(\ref{eq: paraxial equation})), and substituting the transverse derivative operators from Eq.~(\ref{eq: position operator}) leads to the $xz$ propagation direction operator 
\begin{equation}
P_{xz}^{(\pm)}(\phi_x) = \exp\left[\mp i\phi_x\frac{kw_0}{2} \left\{\left(a_x + a_x^{\dag}\right)\left(1+\frac{1}{(kw_0)^2}\left[\left(a_x - a^{\dag}_x\right)^2+ \left(a_y - a^{\dag}_y\right)^2\right]\right)\right\}\right],
\end{equation}
where the exponential is evaluated using the methods introduced in Sec.~\ref{subsec: Taking exponent of surface profile}. Extending this form to more general changes to the propagation direction requires care, but, for the purposes of the analysis in this manuscript, the direction of transverse misalignment defines the $x$ axis, and therefore the propagation direction must lie in the $xz$ plane.

\subsection{Calculating mode angles}
Finally, before effecting the mode rotations of Sec.~\ref{subsec: changing mode angle}, it is often useful to determine the propagation angle of the mode, which can be determined by calculating the expectation value of the angle operator
\begin{subequations}
\begin{align}
    \phi_x^{(\pm)} & = (\mp i/k)\frac{\partial}{\partial x}, \\
    & = (\mp i/(k w_0))(a_x-a^{\dag}_x),
\end{align}
\end{subequations}
which is valid in the paraxial approximation. The eigenstates of this operator are plane waves propagating at angle $\phi_x$ to the $z$ axis in the $xz$ plane. This capability is useful to understand properties of resonant modes for misaligned cavity configurations.

\section{Demonstrating the method}
\label{sec: demonstration}
\subsection{Selecting the mode of interest}
\label{subsec: selecting mode of interest}
The mode mixing method produces a set of cavity eigenmodes $\{\ket{\Psi_{i}}\}$ and corresponding eigenvalues $\{\gamma_i\}$. The important data within these sets are the mode profile and round trip loss of the particular eigenmode that will be used in the application at hand, and therefore a `mode of interest' should be identified. For the majority of applications using spherical cavities, the fundamental mode is more useful than the higher order transverse modes. When the cavity mirrors are transversely misaligned or non-spherical, we expect the propagation angle and central waist of the fundamental mode to change (see Sec~\ref{sec:geometry}). Therefore, for this investigation, the eigenmode chosen is the $\ket{\Psi_i}$ that maximises the overlap $\abs{\bra{\Psi_i}\ket{\Psi_{0, 0}^{G}}}^2$ with the geometrically expected mode denoted $\ket{\Psi_{0, 0}^{G}}$.

The geometric expectation $\ket{\Psi_{0, 0}^{G}}$ has thus far been parameterised through the propagation direction and the central waists in two principal directions, whereas the cavity eigenmodes $\{\ket{\Psi_{i}}\}$ are expressed as coefficients in a basis propagating along the $z$ axis. To find the overlap of the cavity eigenmodes with the expected mode, the cavity eigenmodes were expressed in the same basis as the expected mode by first expanding/contracting in the two transverse directions independently to set the waists, and then rotating the mode in the $xz$ plane to set the propagation direction, according to the methods of Sec.~\ref{subsec: mode transformations}.

\subsection{Comparing to standard methods}
\label{subsec: comparing to standard methods}
\begin{figure*}
\centering\includegraphics[width=0.95\textwidth]{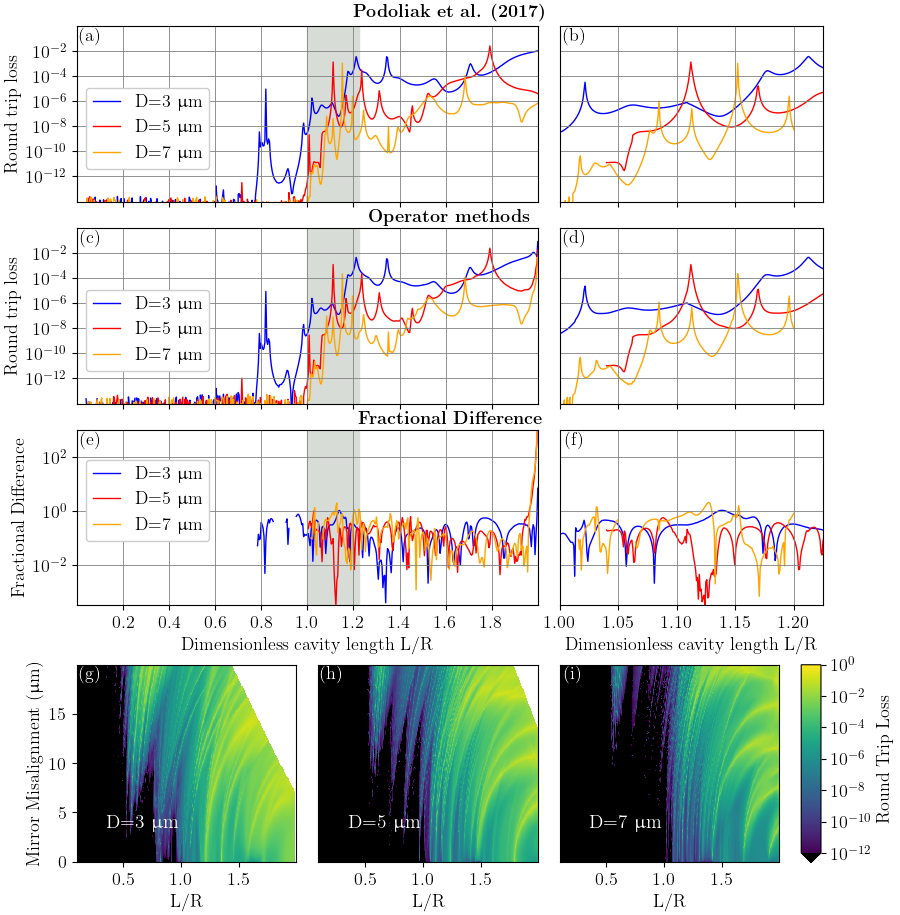}
\caption[Comparison of developed mode mixing methods and results from literature]{Comparison of round-trip loss data for cavities with Gaussian mirrors generated with standard methods (top row) and with the methods described in this manuscript (middle and bottom rows). All mirrors have a central radius of curvature of $R_c=500~\upmu$m, and the interrogation wavelength is 866~nm. a) Round-trip loss data generated using standard techniques for cavities with Gaussian mirrors of three depths as a function of cavity length over the whole stability region and b) in a small region of lengths to the concentric side of the confocal length. Data taken from \cite{Podoliak:17}. The basis used was a Laguerre-Gauss basis up to $n=30$. c) and d) Equivalent round-trip loss data generated using the methods detailed in this manuscript, using a basis of the first 30 even states in both Cartesian directions. e) and f) The difference of the round trip loss predicted by the two methods, expressed as a fraction of the loss predicted by the standard literature method of a) and b). Data not shown where the literature round trip loss is below $10^{-12}$. e)~f) and g) Round-trip loss as a function of length and misalignment for the three mirrors using the translation methods described in Sec.~\ref{subsec: Translating mirror}. Configurations where the region inside one waist of the expected mode (as predicted by the theory of Sec.~\ref{sec:geometry}) would not be fully enclosed within the concave region of the Gaussian mirror are left white. Below round-trip losses of $10^{-12}$, numerical noise becomes significant, and thus these results are left black.}
\label{fig: podoliak comparison}
\end{figure*}

Results obtained using the procedure for constructing mirror matrices using operators (presented in Sec.~\ref{sec: Theory mirror transformation}) were compared with those found in the literature for the case of a Gaussian-shaped mirror (Fig.~\ref{fig: podoliak comparison} a)-d)). The round trip loss was calculated as a function of cavity length for three different Gaussian waist values using both methods. Due to the different calculation bases employed by the methods, the results are not expected to be identical, but should agree up to convergence effects. As shown in Fig.~\ref{fig: podoliak comparison} e) and f) for the vast majority of cases, the methods predict round trip losses with a fractional difference between one hundredth and unity; discrepancies that are practically indiscernible amidst order of magnitude variations described in the data. The exceptions to this are highly concentric configurations, where there is a substantial difference between the losses predicted.

The methods presented for translating the mirror matrices (Sec.~\ref{subsec: Translating mirror}) enable the data generated for aligned configurations to be simply extended to misaligned configurations (Fig.~\ref{fig: podoliak comparison} g)-i)). This capability allows for the round trip loss of cavities with transverse misalignment to be properly simulated, unveiling a rich structure of lossy `bands' in the length-misalignment parameter space that split into multiplets as the misalignment increases. Many of these bands can be traced back to loss peaks in the length scan of the aligned configuration, but some (such as the high loss bands in Fig.~\ref{fig: podoliak comparison}h) appearing to originate from small misalignment at $L/R=1.5~\upmu$m for $D=5~\upmu$m) cannot. This implies that residual misalignment introduces mechanisms of loss that do not feature for perfectly aligned cavities. A detailed discussion of the physics of misaligned cavities is beyond the scope of this paper, but will instead be the subject of a future publication.

\section{Conclusion}
\label{sec: Conclusion}

We have developed methods to calculate the modes of cavities with non-spherical and transversely misaligned mirrors. We used a classical ray model to predict the mode axis and central waist of the resonant mode of a misaligned cavity, using these results to understand the output of a more complete mode mixing method. This method is inspired by existing techniques that exploit well-known operator forms and transformation matrices to model mode mixing in cavities with different mirror profiles, and to simply extend these models to include mirror misalignment.

The theory introduced in this paper is applicable to a variety of mode-mixing scenarios. Firstly, for particular mirror shapes where the deviation from the ideal parabolic profile can be expressed as a sum of polynomials in the transverse coordinates, or as a Gaussian function, the mode mixing matrix is calculated using analytical results and a matrix exponential, removing the need for any overlap integrals of the basis functions with the mirror profile to be taken. Secondly, once the mirror matrix has been obtained, the mirror can be translated using operators (which also do not require integrals to be calculated). This allows for the cavity mode structure under transverse misalignment to be determined in a simple manner and, in our experience, more quickly than with conventional techniques.  

We anticipate the methods developed in this work will find application in the simulation of optical resonators with non-spherical mirrors, particularly for cases where the transverse misalignment of the mirrors is not negligible. An analysis of cavities with Guassian-shaped mirrors utilising the methods of this work will be the subject of a future publication.

\begin{acknowledgments}
This work was funded by the UK Engineering and Physical Sciences Research Council Hub in Quantum Computing and Simulation (EP/T001062/1) and the European Union Quantum Technology Flagship Project AQTION (No. 820495). The authors would like to acknowledge the use of the University of Oxford Advanced Research Computing (ARC) facility in carrying out this work. http://dx.doi.org/10.5281/zenodo.22558. Data underlying the results presented in this paper are available in Ref. \textbf{DOI added on acceptance}. The code that generated the data may be obtained from the authors at reasonable request.
\end{acknowledgments}

\bibliographystyle{unsrt}

\appendix

\section{Derivation of Operators in Hermite Gauss Basis}
\label{app: derivation_of_HG_operators}
To derive the operator forms of $x$ and $\frac{\partial}{\partial x}$, we start by comparing the mode amplitude of the basis states introduced in Eq.~(\ref{eq: Hermite Gauss mode})

\begin{equation}
\begin{aligned}
    u^{(\pm)}_{n_x,n_y}\left(x,y,z\right)  =  a(z) H_{n_x}\left(\frac{\sqrt{2}x}{w(z)}\right)H_{n_y} & \left(\frac{\sqrt{2}y}{w(z)}\right) \\
\exp\left[-\frac{x^2+y^2}{w(z)^2}\right] & \exp\left[\mp ik\frac{x^2+y^2}{2R_u(z)}\right]\exp\left[\pm i(n_x+n_y+1)\Psi_G\right],
\end{aligned}
\label{eqn: HG_mode_definition_appendix}
\end{equation}
with the mode of the quantum harmonic oscillator of mass $m$ and resonant frequency $\Omega$ 

\begin{equation}
\psi^{\mathrm{HO}}_{n_x,n_y}(x,y) = H_{n_x}\left(\sqrt{\frac{m\Omega}{\hbar}}x\right)H_{n_y}\left(\sqrt{\frac{m\Omega}{\hbar}}y\right)\exp\left[-\frac{m\Omega(x^2+y^2)}{2\hbar}\right].
\label{SHO wavefunctions}
\end{equation}

The quantum harmonic oscillator has operators
\begin{subequations}
    \begin{align}
        x^{\mathrm{HO}} & = \sqrt{\frac{\hbar}{2m\Omega}}(a_x^{\mathrm{HO}}+(a_x^{\mathrm{HO}})^{\dag}), \\
\frac{\partial}{\partial x}^{\mathrm{HO}} & = \sqrt{\frac{m\Omega}{2\hbar}}(a_x^{\mathrm{HO}}-(a_x^{\mathrm{HO}})^{\dag}),
    \end{align}
    \label{eq: SHO operators}
\end{subequations}
shown in terms of the harmonic annihilation operator $a_x^{\mathrm{HO}}$ \cite{Schwinger:01}. In the case that the parameters of the harmonic oscillator and Gaussian mode are related by $m\Omega/2\hbar=1/w^2$, the respective wavefunctions are related by
\begin{equation}
u^{\pm}_{n_x,n_y}\left(x,y\right) = \psi(x,y)^{\mathrm{HO}}_{n_x,n_y}\exp\left[\mp ik\frac{x^2+y^2}{2R}\right]\exp\left[\pm i(n_x+n_y+1)\Psi_G\right]. 
\end{equation}
Therefore, the $x$ operator in the cavity mode basis set can be found in terms of the $x$ operator in the harmonic oscillator basis

\begin{subequations}
\begin{align}
x^{(\pm)}_{n_x',n_y',n_x,n_y} & = \int_S u^{\pm*}_{n_x', n_y'}(x,y) x u^{\pm}_{n_x,n_y}(x,y) \, dxdy, \\
x^{(\pm)}_{n_x',n_y',n_x,n_y} & = \exp\left[\pm i\Psi_G(n_x+n_y-n_x'-n_y')\right]\int_S \psi^{* \mathrm{HO}}_{n_x',n_y'} (x,y)x \psi^{\mathrm{HO}}_{n_x,n_y}(x,y) \, dxdy, \\
x^{(\pm)}_{n_x',n_y',n_x,n_y} & = x^{\mathrm{HO}}_{n_x',n_y',n_x,n_y}\exp\left[\pm i\Psi_G(n_x+n_y-n_x'-n_y')\right].
\end{align}
\label{SHO wavefuntion analogy x explicit}
\end{subequations}
The analogy between the wavefunctions then leads to
\begin{subequations}
\begin{align}
x^{(\pm)} & = (U_G^{(\pm)})^{\dag}\frac{1}{2}w(a_x+a^{\dag}_x)U_G^{(\pm)},\\
(U_G)^{(\pm)}_{n_x',n_y',n_x,n_y} & = \delta_{n_x', n_x}\delta_{n_y',n_y}\exp\left[\pm i\Psi_G(n_x+n_y+1)\right], 
\end{align}
\end{subequations}
where $a_x$ is the annihilation operator in the $x$-direction for the mode functions $u^{(\pm)}_{n_x,n_y}\left(x,y,z\right)$, which acts equivalently to the $a_x^{\mathrm{HO}}$ operator on the harmonic oscillator wavefunctions.
A similar approach can be used for the $\frac{\partial}{\partial x}$ operator
\begin{subequations}
\begin{align}
\frac{\partial}{\partial x}^{(\pm)}_{n_x',n_y',n_x,n_y} & = \int_S u^{\pm*}_{n_x',n_y'}(x,y) \frac{\partial}{\partial x} u^{\pm}_{n_x,n_y}(x,y) \, dxdy, \\
\frac{\partial}{\partial x}^{(\pm)}_{n_x',n_y',n_x,n_y}  & = \exp \left[i\Psi_G(n_x+n_y-n_x'-n_y')\right] \times \nonumber\\  \int_S \left( \psi^{*\mathrm{HO}}_{n_x',n_y'}(x,y) \frac{\partial}{\partial x}  \psi^{\mathrm{HO}}_{n_x,n_y}(x,y)\right) &+\left( \psi^{*\mathrm{HO}}_{n_x',n_y'}(x,y) \left(\mp ik\frac{x}{R}\right)\psi^{\mathrm{HO}}_{n_x,n_y}(x,y)  \right) \, dxdy, \\
\frac{\partial}{\partial x}^{(\pm)} & = (U_G^{(\pm)})^{\dag}\left(\frac{\partial}{\partial x}^{\mathrm{HO}} \mp \frac{ik}{R}x^{\mathrm{HO}}\right)U_G^{(\pm)},
\end{align}
\label{SHO wavefuntion analogy ddx explicit}
\end{subequations}
resulting in the expression
\begin{equation}
\frac{\partial}{\partial x}=(U_G^{(\pm)})^{\dag}\left(\frac{1}{w}(a_x-a^{\dag}_x) \mp i\frac{wk}{2R}(a_x+a^{\dag}_x)\right)U_G^{(\pm)},
\end{equation}
which can be converted algebraically to the more convenient form
\begin{equation}
\frac{\partial}{\partial x}=\left(\frac{1}{w_0}(a_x-a^{\dag}_x)\right).
\end{equation}

\section{Finding the Gaussian Profile Matrix}
\label{app: finding_gaussian_profile_matrix}
The formula for the Gaussian profile surface matrix in the Hermite-Gauss basis (Eq.~(\ref{eq: Gaussian surface operator})) is calculated following the method of appendix A of \cite{Varro:22}. To evaluate a unit-depth one-dimensional Gaussian as a matrix, we start by expanding using the transverse coordinate operator of Eq.~(\ref{eq: position operator})
\begin{subequations}
\begin{align}
\exp\left[-\frac{x^2}{w_e^2}\right] & = (U_G^{(\pm)})^{\dag} \exp\left[\chi \left(\frac{1}{2}\left(a^{\dag}\right)^2+\frac{1}{2}\left(a^{\dag}\right)^2+\frac{1}{2}\left(aa^{\dag}+a^{\dag}a\right)\right)\right] U_G^{(\pm)}, \\
& = (U_G^{(\pm)})^{\dag} \exp\left[\chi \left(K_{+}+K_{-}+2K_0\right)\right] U_G^{(\pm)}, \\
\chi & = -\frac{w^2}{2w_e^2},
\end{align}
\end{subequations}
where $K_+$, $K_-$ and $K_0$ are $(1/2)\left(a^{\dag}\right)^2$, $(1/4)\left(a^{\dag}\right)^2$, and $(1/2)\left(aa^{\dag}+a^{\dag}a\right)$ respectively, and the annihilation operator $a$ represents $a_x$ ($a_y$) for the $x$ ($y$) directed Gaussian function. Now use that $K_+$, $K_-$ and $K_0$ have the same commutation relations as $-\sigma_+$, $\sigma_-$ and $(1/2)\sigma_3$, where
\begin{equation}
-\sigma_+  = \begin{pmatrix}
0 & -1\\
0 & 0 
\end{pmatrix}, \quad
\sigma_-  = \begin{pmatrix}
0 & 0\\
1 & 0 
\end{pmatrix}, \quad
\frac{1}{2}\sigma_3  = \frac{1}{2}\begin{pmatrix}
1 & 0\\
0 & -1
\end{pmatrix}.
\end{equation}

Next, we equate the coefficients of the exponent and normal-ordered exponents of the 2-dimensional matrices, where the normal form has coefficients $\zeta$, $\zeta '$ and $\eta$
\begin{equation}
    \exp\left[\zeta \left(-\sigma_+\right)\right] \exp\left[-\eta \left(\sigma_3\right)\right] \exp\left[\zeta '  \left(\sigma_-\right)\right] = \exp\left[\chi  \left(-\sigma_+\right) + \chi  \left(\sigma_-\right) + 2\chi  \left(\sigma_3\right)\right].
\end{equation}
Expanding the two sides of this equation gives
\begin{equation}
\zeta = \zeta ' =  \frac{\chi}{1-\chi}, \quad
\eta = 2\ln \left(1-\chi\right).
\end{equation}
The 2-dimensional matrices are substituted back for creation and annihilation operators to obtain
\begin{equation}
\begin{aligned}
    \exp\left[\chi \left(K_{+}+K_{-}+2K_0\right)\right] = &  \\  \exp\left[\frac{\chi}{1-\chi}\frac{1}{2}\left(a^{\dag}\right)^2\right] &\exp\left(-2\ln \left(1-\chi\right)\frac{1}{4}\left(a a^{\dag} + a^{\dag}a \right)\right)\exp\left(\frac{\chi}{1-\chi}\frac{1}{2}\left(a\right)^2\right). 
\end{aligned}
\end{equation}
The normal operator form can be evaluated simply in the Hermite Gauss basis to obtain the result quoted in Sec.~\ref{subsec: calculate Gaussian surface profile}:
\begin{subequations}
\begin{align}
\exp(-\frac{x^2}{w_e^2})^{(\pm)}_{m',m} & = \nonumber\\
(U_G^{(\pm)})^{\dag} \left(1-\chi\right)^{-(\frac{m'+m+1}{2})} & \left(\frac{\chi}{2}\right)^{\frac{m'-m}{2}}\sqrt{m'!m!}\sum_{k=0}^{[\frac{m}{2}]} \frac{\left(\frac{\chi^2}{4}\right)^k}{\left(\frac{m'-m}{2}+k\right)!k!\left(m-2k\right)!} U_G^{(\pm)}, \\
\chi & =  -\frac{1}{2}\frac{w\left(z\right)^2}{w_e^2}. 
\end{align}
\end{subequations}

\end{document}